\documentstyle[11pt]{article}
\begin{document}
\vskip 0.2in
\begin{center}
\vskip 0.1in
\Large{\bf
LARGE MASS DIPHOTONS FROM}\\
\vskip 0.1in
\Large{\bf RELATIVISTIC HEAVY ION COLLISIONS}\\
\vskip 0.2in
\large{\bf Sourav Sarkar$^a$,
Dinesh Kumar Srivastava$^a$, Bikash Sinha$^{a,b}$,}
\vskip 0.1in
\large{\bf Pradip Kumar Roy$^a$, Subhasis Chattopadhyay$^a$, and Dipali Pal$^a$} 
\vskip 0.2in
\small\it {$^a$Variable Energy Cyclotron Centre,
     1/AF Bidhan Nagar, Calcutta 700 064,
     India}
\vskip 0.1in
\small\it{$^b$Saha Institute of Nuclear Physics,
1/AF Bidhan Nagar, Calcutta 700 064, India}
\end{center}

\renewcommand{\baselinestretch}{1.5} 
\parindent=20pt

\vskip 0.1in
\begin{center}
\bf {Abstract}\\
\end{center}
We evaluate the production of large mass diphotons from quark 
annihilation at BNL RHIC and CERN LHC energies from  central
collisions of gold nuclei.
The collision is assumed to lead to either 
a thermally and chemically equilibrated quark gluon plasma, 
or a free-streaming quark gluon gas having an identical initial
entropy, or a chemically equilibrating quark gluon system,
with the same entropy at $T=T_c$. 
We also obtain an estimate of hard photon pairs
from initial state quark annihilation and find that the
thermal production dominates the yield up to $M \approx $ 4 GeV
at RHIC, and up to 6 GeV at LHC.
A simulation study of decay versus thermal diphotons is presented.
\vskip 0.25in

\vskip 30pt


\vskip 0.5in

\newpage
\addtolength{\baselineskip}{0.6 \baselineskip}

Relativistic heavy ion collisions are expected to lead to a
confirmation of the QCD phase transition from hadronic matter to
quark matter and an ephemeral formation of quark
gluon plasma (QGP).
Analysis of the S and Pb induced collisions at the CERN SPS
have brought the search for QGP to an interesting stage.
 Many of the proposed signatures
of QGP have been observed, though, most of them can also be explained 
to some extent in a purely hadronic picture. 
It is expected that the issue will be clearly resolved when
we have results from BNL RHIC and CERN LHC, as much larger initial
temperatures are likely to be attained there.
 
 Dileptons have long been considered as excellent probes of
the early stages of relativistic heavy ion collisions.
Large mass dileptons are likely to have their origin
in the hot and dense stage of QGP and their detection would provide 
valuable information about this exotic state of matter. 
 We suggest that an experimental detection of large mass
diphotons can possibly provide a valuable confirmation of the results 
obtained from the measurement of dileptons. The theoretical
understanding of the basic rates as well as the evolution of the
interacting system has improved considerably since the early
studies~\cite{early} in this direction, necessitating a re-evaluation
of the importance of diphoton measurements. Thus, we have utilized the
thermal mass of the annihilating quarks while evaluating the rates, which is
more appropriate for  quarks in a plasma.
 We have also considered evolution mechanisms
which encompass the entire range of possible scenarios describing the
expansion of the QGP. We give, as far as we know, the first estimate of
hard QCD diphotons from the colliding nuclei. We
also present a feasibility study of these measurements by performing a 
simulation of the likely measurements.

  Large mass diphotons from the QGP are produced from  quark -
antiquark annihilation. The cross section is given by~\cite{early}:
\newpage

\begin{eqnarray}
\sigma_{q\bar{q}}^{\gamma\gamma}(M)&=&2\pi\alpha^2\,N_c\,(2S+1)^2\,
\sum_q\frac{e_q^4}{M^2-4m_q^2}\nonumber\\
& &\left[\left[1+\frac{4m_q^2}{M^2}-\frac{8m_q^4}{M^4} \right]
\,\ln\left\{\frac{M^2}{2m_q^2}
 \left[1+\left[1-\frac{4m_q^2}{M^2}\right]^{1/2}\right]-1\right\}
\right.\nonumber\\
& &\left.-\left[1+\frac{4m_q^2}{M^2}\right]\left[1-\frac{4m_q^2}{M^2}\right]
^{1/2}\right],\\
&\stackrel{M >> m_q}{=}&2\pi\alpha^2\,N_c\,(2S+1)^2\,
\sum_q\frac{e_q^4}{M^2}\ln\left\{\frac{M^2}{2.718m_q^2}\right\}.
\end{eqnarray}
In the above $N_{\mathrm {c}}=3$, $S=1/2$ and $e_{\mathrm {q}}$ 
is the charge of the quark.  This  cross section 
diverges as $m_{\mathrm {q}}~\rightarrow~0$.
The early works on diphotons~\cite{early} mostly used $m_q=$ 5 MeV,
which grossly overestimates the rates. We argue that as
quarks in a heat bath acquire a thermal mass~\cite{Braat}
given by $m_{\mathrm {th}}=\sqrt{(2\pi\alpha_s/3)}\,T$, this 
singularity can be regularized by making the natural choice,
$m_{\mathrm {q}}=m_{\mathrm {th}}$.  One can 
easily check that choosing $m_q=$ 5 MeV, instead, will increase the
basic cross-section by a factor of $\sim$ 3--5.
Increasing $m_q$ from $m_{\mathrm {th}}$ to $2 m_{\mathrm{th}}$, say,
will decrease the cross-section by less than 50\%.
 We shall see later that the explicit appearance of $T$ in 
the basic cross-section here provides us with an additional probe for
 the evolution of the temperature~\cite{com}.

  As remarked earlier, we investigate the production of thermal diphotons
 in three
different scenarios, which essentially encompass the entire range of
evolution dynamics one can imagine and thus give us the upper and the
lower limits of the diphoton yield.
As a first step, we consider the formation of a QGP in a
thermodynamic (thermal as well as chemical) equilibrium at an initial 
time $\tau_0$ and initial temperature $T_0$ in a central collision of
two gold nuclei. This case has been studied in great 
detail in the literature. One can use the condition of isentropic expansion 
to relate the initial conditions to the particle multiplicity
density $(dN/dy)$~\cite{HK}; 
\begin{equation}
T_0^3\,\tau_0=\frac{2\pi^4}{45\zeta(3)\pi R^2 4a_Q}\frac{dN}{dy}.
\end{equation}
For a  QGP  consisting of u, d, and s
quarks, and gluons, $a_Q=47.5\pi^2/90$. $R$ is the
initial transverse dimension of the system.
Assuming a rapid thermalization, one may get an upper limit on 
the temperature by taking $\tau_0=1/3T_0$. 
Taking $dN/dy=$ 1735 for central collision of Au nuclei  at RHIC 
we get an initial temperature of 478 MeV, while the corresponding numbers at
LHC are 5624 and 860 MeV respectively~\cite{KMS}.  As we are interested 
in large mass photon pairs, we include only the emission from the  
QGP phase~\cite{dks}, i.e., till the system reaches the phase transition 
temperature $T_c$. 
We further neglect the transverse expansion which is known to affect
the large invariant mass distribution only marginally~\cite{Larry}.
The proper time $\tau_q$, when the temperature drops to $T_c$, is
obtained from the Bjorken relation $\tau_q=\tau_0{(T_0/T_c)}^3$~\cite{Bj}.
The invariant mass spectrum for diphotons will then be,
\begin{equation}
\frac{dN}{dM^2dy}=\frac{\pi R^2\,M^3}{2(2\pi)^4}
{\int_{\tau_0}}^{\tau_q} \sigma(M,T)(1-4m_q^2/M^2)T(\tau)K_1(M/T)
\tau\, d\tau.
\end{equation}

At the other extreme, one may imagine that the system is formed at a proper time
$\tau_0$ with a temperature $T_0$ after which the constituents 
free-stream away from the collision zone.
One can evaluate the yield of diphotons from such a gas, in a manner
similar to Ref.~\cite{KMS}, as
\begin{equation}
\frac{dN}{dM^2dy}=\pi
R^2\,\frac{M^2\sigma(M,T_0)T_0^2{\tau}_0^2}{2^7\pi^3}\,\ln\left(
\frac{R}{\tau_0}\right)\,
\,\exp\,(-M/T_0)\left(1+2.19\frac{T_0}{M}\right).
\end{equation}

It is quite likely that the quark gluon system may 
be neither in thermal
equilibrium nor in chemical equilibrium when produced initially. 
Even though a thermal (kinetic) equilibrium may be attained quickly
~\cite{Klaus}, the system may still be away from chemical equilibrium,
 which may evolve
with passage of time due to a number of partonic reactions.
Bir\'{o} {\it et al.}~\cite{Biro}, for example,
have studied the evolution of chemical equilibration
with initial conditions determined from the HIJING model~\cite{Wang}.
In order to have a meaningful comparison with the two descriptions
above, we use the time variations of quark fugacity and 
temperature obtained by Strickland~\cite{Strick},
 with the stipulation that the particle multiplicity density at 
the end of the QGP phase is equal to the values of $dN/dy$ used here.
Now the diphoton spectrum is obtained from an expression similar to Eq.~4
with the additional introduction of the quark fugacity ($\lambda_q$),
taken from the Fig.~1 of Ref.~\cite{Strick}.

In Fig.~1 we give our results for the
invariant mass distribution of the large mass photon pairs
for RHIC energies. A number of observations are in order here. While the 
slope of the free-streaming description is essentially determined by $T_0$
(see Eq.~5), the temperature drops from 
$T_0$ to $T_c$ in the hydrodynamic
description, thus leading to an effectively steeper slope.
Further, the yield for the equilibrating plasma is smaller due to 
the appearance of the square of the quark-fugacity in the expression and a more rapid cooling.  Similar results are seen for the LHC energies as well
 (see Fig.~2).

 We must add here that the results for the equilibrated
hydrodynamic evolution and those for the free-streaming are obtained
by using $\tau_0=1/3T_0$ (see above), while those for the non-equilibrium
hydrodynamics have $\tau_0=0.3 $ fm/$c$. One may get yet another estimate for
these measurements by starting the equilibrated hydrodynamic evolution
from $\tau_0=0.3$ fm/$c$ (see Figs.~1 \& 2).
Now the results for small $M$ get quite similar
to that for the non-equilibrium hydrodynamics. This looks surprising at first.
Recall that our evolution scenarios are tailored to lead to 
a  given final entropy (multiplicity). This ensures that during the later
stages the conditions of the equilibrium and the 
non-equilibrium hydrodynamic evolution scenarios are nearly identical,
leading to the similarity of these predictions at low $M$.

The diphoton production from the hard QCD annihilation
 of quarks in the colliding nuclei at $y=0$ is given by;
\begin{eqnarray}
\frac{dN}{dM^2dy}&=&\frac{2\pi \alpha^2}{sM^2N_c}T_{AB}\left[
\ln\left(\frac{M^2-p_c^2}{p_c^2}\right)-\left(1-2p_c^2/M^2\right)
\right]\nonumber\\
& &\times \sum_q e_q^4
\left[q^{N_p}(x,M^2)\bar q^{N_t}(x,M^2)+\bar q^{N_p}(x,M^2)
q^{N_t}(x,M^2)\right],
\end{eqnarray}
where $x=M/\sqrt s$ and $\sqrt s$ is the c.m. energy of the
colliding nucleons. $N_p$ and $N_t$ denote nucleons in the projectile
and target respectively. We have introduced an arbitrary cut-off on the 
momentum  transfer $p_c$ (=2 GeV), so that pQCD remains applicable for
 such cases and used the MRSD$-^\prime$ set of 
parametrizations~\cite{MRSD} for the quark structure functions. 
$T_{\mathrm {AB}}$ (= 293.4 fm$^{-2}$) is the nuclear thickness function
at zero impact parameter, corresponding to central Au$+$Au collisions.
The result of this analysis for the RHIC and LHC energies are 
also given in Figs.~1 and 2, respectively.
We find that thermal pairs dominate the
spectrum up to 4 GeV at RHIC and up to 6 GeV at LHC. 

 What additional information will large mass diphotons provide
which we shall not already have from the measurement of dileptons?
To explore this, we have evaluated the invariant mass distribution
for dileptons as well as diphotons produced from a  
 thermally and chemically equilibrated QGP, which
expands and  cools to a temperature $T_c$ (160 MeV), goes into a mixed phase
where a phase transition occurs, and when all of the
QGP has converted into hadrons, again expands and cools till 
freeze-out temperature $T_f$ (100 MeV) is reached. 
We consider a boost invariant longitudinal expansion without and with,
a cylindrically symmetric, transverse expansion. We include the
annihilation of pions in the hadronic matter and annihilation of quarks
in the QGP as sources of dileptons~\cite{Larry} and diphotons~\cite{early,dks}.

In Fig.~3 we have plotted the ratio of diphotons to
dileptons as a function of $M$ at RHIC and LHC energies. 
Firstly, we note that the structure 
seen in this ratio has its origin in the pion form-factor leading to the
dileptons. The minimum at $ M \approx $ 0.8 GeV comes from the $\rho$ peak
in the pion form-factor and the rapid rise there-after reflects the
sharp decrease in this. The maximum around 1.4 GeV corresponds to the
region where the (structure-less) quark annihilation process starts
dominating. The differences in the $1+1$ and the $3+1$ dimensional cases are
mostly confined to lower masses, which have their origin in the late
stages of the evolution when the transverse expansion of the system
is truly large. 

It is easy to verify that {\it if} we ignore the $T$ 
dependence of the thermal mass of the quarks appearing in Eq.~1, a 
universal curve would be obtained for this ratio at large masses at
 all energies. 
The temperatures likely to be attained at LHC are larger than those
at RHIC. Thus, in a purely longitudinal expansion these ratios at
larger masses are smaller at LHC (recall the factor 
$\sim\,\ln(M^2/2.7m_q^2)$ in expression for cross-section given by Eq.~2).
The transverse expansion of the system leads to a more rapid cooling of
the plasma and thus higher temperatures are retained only for a shorter
duration.  This will further increase these ratios, as indeed we see
 from Fig.~3.  Thus we note that large mass
diphotons sensitively reveal the details of the dynamics
of the evolution of the plasma, especially when studied together with 
dileptons. The predicted variation, if observed, will also
confirm our understanding of the temperature dependence of the
thermal mass of quarks. 

Will it be possible at all to isolate diphotons from the huge background of
photons coming from the decay of $\pi^0$ mesons? 
It is quite clear that the background of decay photons can be
arbitrarily reduced by choosing a large enough $p_T$.
 The search for diphotons will have to start with an accurate reconstruction
 of the decay photons  so that any new source of photon pairs could be
identified.  We utilize the method of mixed-event analysis pioneered by the
WA80 collaboration.  We illustrate the situation at LHC energies and
 approximate the $p_T$ distribution of pions $\sim~ \exp(-p_T/T)$, with
 $T \simeq$ 300 MeV.  The invariant mass spectrum for the 
photon-pairs from the decay of these pions in a sample of 
ten million central events
 is shown in Fig.~4(a). Similarly the invariant mass spectrum for the
mixed events was also generated. This constitutes the back ground (see, e.g.,
Ref.~\cite{mix}).
A subtraction of the later from the invariant mass spectrum 
for the real events is shown in Fig.~4(b).
The $\pi^0$ peak is seen very clearly. The  diphotons
from our calculations are seen to stand out clearly against this
background. This simulation holds out the hope that diphotons could be
isolated if high statistics data are available. It is important to note that
the structure of the subtracted back-ground is quite
different from the structure of the diphoton signal. Very similar
results were obtained by fitting the background (e.g., in Fig.~4a) to a
polynomial and subtracting the fitted spectrum from the invariant mass
distribution~\cite{mix}.

In brief, we have evaluated the large mass
 invariant mass spectrum of thermal
diphotons for a number of plausible scenarios in relativistic heavy ion 
collisions. An estimate of prompt diphotons is also made. It is shown
that a comparison of diphotons and dileptons from heavy ion collisions can
lead to an interesting confirmation of the temperature dependence of the thermal
mass of quarks. Finally a simulation of decay  versus thermal diphoton
production at LHC energies is performed which shows that such measurements
could indeed be feasible, once high statistics data are available.

We are grateful to Patrick Aurenche for useful communications 
in connection
with the evaluation of prompt photon pairs. We are also grateful to
Michael Strickland for providing us tabulated values of
$T(\tau)$ and $\lambda_q(\tau)$~\cite{Strick}. We thank Terry Awes and
Itzhak Tserruya for useful suggestions in connection with the simulation study
carried out here.         

\vskip 0.3in


\newpage
\section*{Figure Captions:}

\noindent Figure 1: Invariant mass distribution of photon pairs
from $q\bar q$ annihilation for the Au$+$Au system at RHIC energies in
a fully equilibrated QGP (Eq. Hydro), a free-streaming gas of quarks and
gluons, and a chemically equilibrating (Noneq. Hydro) quark gluon system.
Invariant mass distribution of  hard photon
pairs (Hard) from initial state $q\bar q$ annihilation for
momentum transfer greater than 2 GeV is also shown.

\vskip 0.25 in

\noindent Figure 2: Same as Fig.~1 for LHC energies.

\vskip 0.25 in

\noindent Figure 3: Ratio of diphoton and dilepton invariant
mass distribution for a hadronizing QGP at RHIC and LHC energies. 
Sources are $q\bar q
\rightarrow \gamma\gamma$ and $q\bar q \rightarrow \mu^+\mu^-$
in the QGP, and $\pi^+\pi^- \rightarrow \gamma\gamma$ and
$\pi^+\pi^- \rightarrow \mu^+\mu^-$ in the hadronic phase.
The solid and dashed lines indicate calculations with and without
transverse expansion. 

\vskip 0.25 in

\noindent Figure 4: (a) Distribution of invariant mass generated from
photon pairs of the same events at LHC energies. (b) Invariant mass
distribution of thermal diphotons and the mixed-event subtracted background
of photon pairs. The difference is normalized so that the integral from
$M=$ 0.2 to 3.5 GeV vanishes. The hatched bar shows the $\pi^0$ peak.
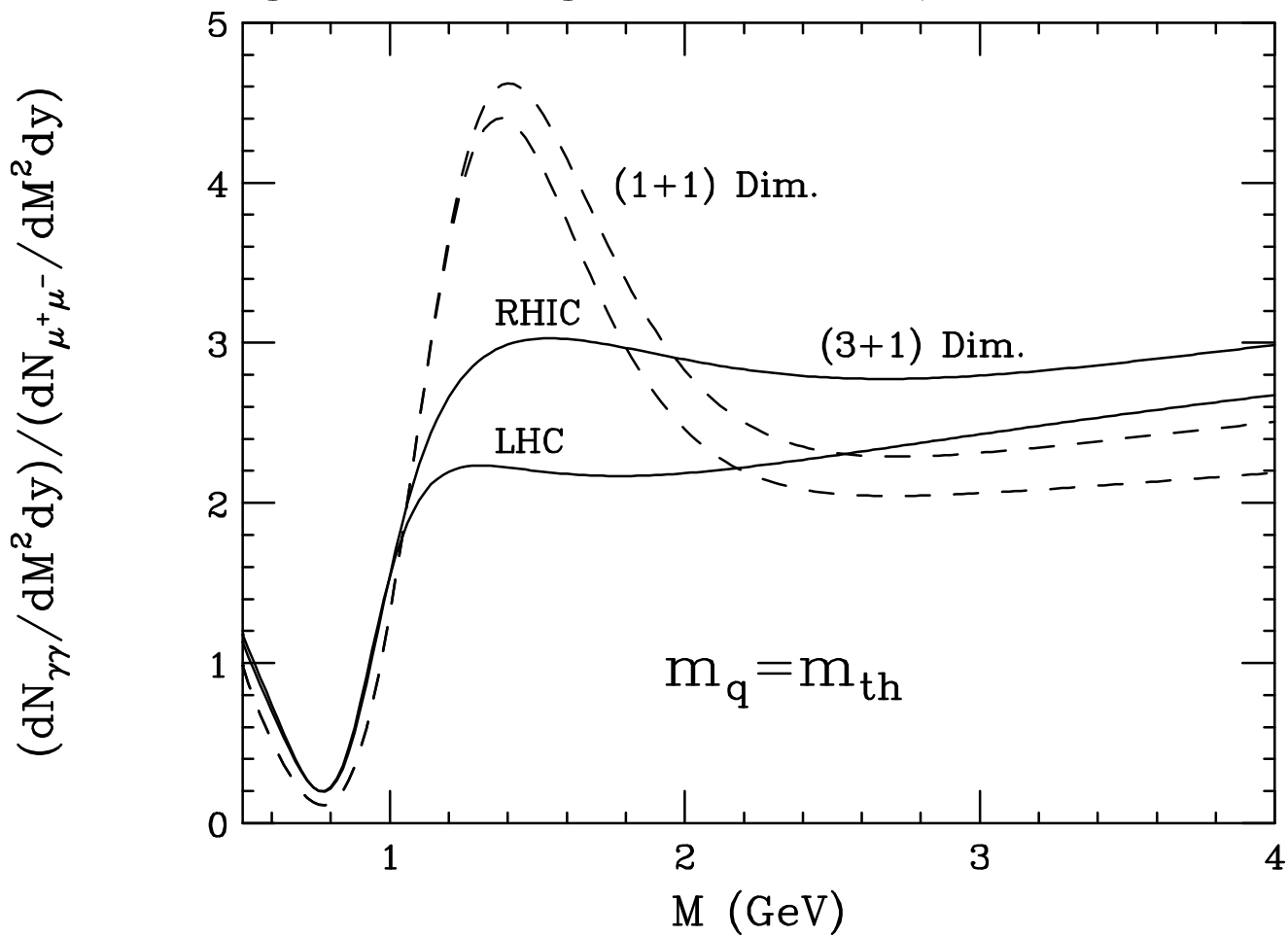
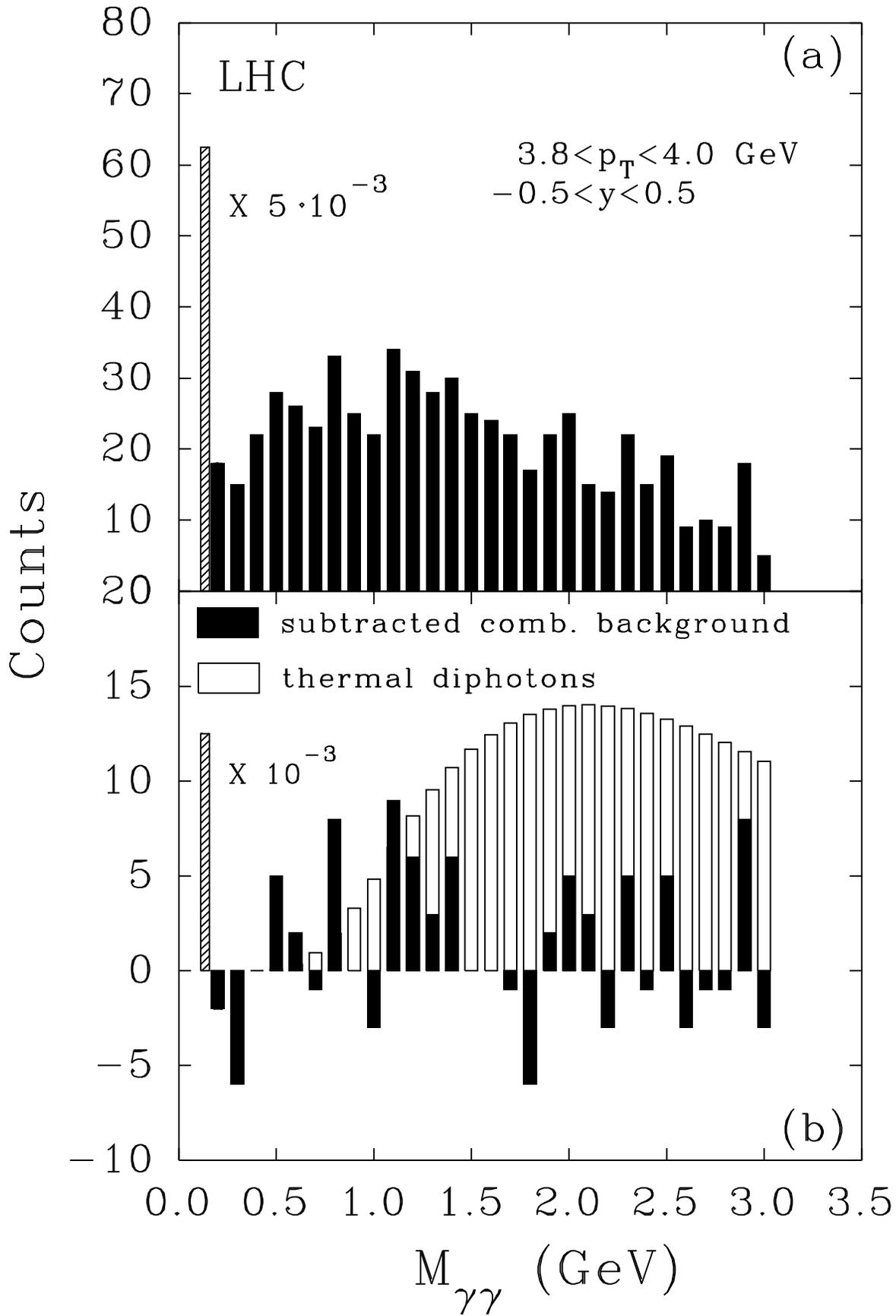
\end{document}